\documentclass[aps,twocolumn,superscriptaddress,final]{revtex4}

\usepackage{graphicx}
\usepackage{amsmath}
\usepackage{amsfonts}

\newcommand{\transpose}[1]{#1^\mathrm{T}}
\newcommand{\mean}[1]{\langle #1\rangle}
\newcommand{\chis}[1]{\chi_{\scriptscriptstyle #1}}
\newcommand{\minimum}{\textrm{\scriptsize{min}}}

\begin{document}

\title{Tight bound on coherent states quantum key distribution with heterodyne detection}

\author{J\'er\^ome Lodewyck}
\affiliation{Thales Research and Technologies, RD 128,
 91767 Palaiseau Cedex, France}
\affiliation{Laboratoire Charles Fabry de l'Institut d'Optique, CNRS UMR 8501,\\
Campus Universitaire, b\^atiment 503, 91403 Orsay Cedex, France}
\author{Philippe Grangier}
\affiliation{Laboratoire Charles Fabry de l'Institut d'Optique, CNRS UMR 8501,\\
Campus Universitaire, b\^atiment 503, 91403 Orsay Cedex, France}

\begin{abstract}
	We propose a new upper bound for the eavesdropper's information in the direct and reverse reconciliated coherent states quantum key distribution protocols with heterodyne detection. This bound is derived by maximizing the leaked information over the symplectic group of transformations that spans every physical Gaussian attack on individual pulses. We exhibit four different attacks that reach this bound, which shows that this bound is tight. Finally, we compare the secret key rate obtained with this new bound to the homodyne rate.
\end{abstract}

\maketitle

\section{Introduction}
\label{section:intro}

	Continuous variable quantum key distribution (CVQKD) is an alternative to single photon ``discrete" QKD that encodes key information in variables with a continuum of degrees of freedom. Such variables include the quadratures $X$ and $P$ of a mode of the electromagnetic field. A CVQKD protocol using these quadratures has been introduced in~\cite{Fred:nature}. It consists in sending a train of pulsed coherent states modulated in $X$ and $P$ with a Gaussian distribution (Alice's module), and in quadrature measurements with an homodyne detection upon reception (Bob's module). Then, Bob's continuous data are used as the basis for constructing a secret binary encryption key, in a classical information process called ``reverse reconciliation" (RR). The security of the RRCVQKD homodyne protocol has first been proven against individual Gaussian attacks~\cite{Fred:nature}, and later extended to individual or finite-size non-Gaussian attacks~\cite{Fred:ng}. More recently, new security proofs covering collective, Gaussian and non-Gaussian attacks~\cite{Fred:col, Raul:col} have appeared.

\begin{figure}
\begin{center}
	\includegraphics[width=0.8\columnwidth]{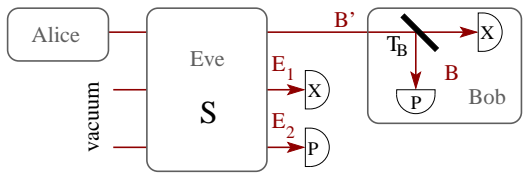}
	\caption{Heterodyne protocol. Bob measure both quadratures $X$ and $P$ of the incoming mode $B'$. A generic eavesdropping strategy involves a transformation $S$ on Alice's mode and two vacuum ancillary modes.}
	\label{fig:heterogeneral}
\end{center}
\end{figure}

	In the homodyne protocol~\cite{Fred:nature}, Bob randomly chooses to measure \emph{either} $X$ or $P$, and then announces his choice. Another possible approach, proposed in~\cite{Weedbrook:noswitch1} by Weedbrook and co-workers, is that Bob measures \emph{both} $X$ and $P$ quadratures of each incoming coherent state, by separating them on a 50-50 beam-splitter $T_B$. This detection, called ``heterodyne'' is represented in Fig.~\ref{fig:heterogeneral}. Notably, in this protocol, Bob does not need to randomly switch his measurement basis.

	In~\cite{Weedbrook:noswitch1}, the authors proposed a bound on the information acquired by the eavesdropper (Eve) in the heterodyne protocol, under the hypothesis of individual Gaussian attacks. They also considered a possible attack based on feed-forward (see below for details), that they however found to be suboptimal with respect to their proposed bound. Therefore, a gap remained between the apparently most stringent bound and the best known attack, which is surprising in the simple scenario of individual Gaussian attacks. Later on, in~\cite{Weedbrook:noswitch2}, the same authors conjectured that their proposed attack is indeed optimal, and so that tighter bounds should apply to Eve's information. However, no such tighter bound was proposed so far.

	In this article, we propose a new bound for individual Gaussian attacks on the CVQKD heterodyne protocols, tighter than the bound of~\cite{Weedbrook:noswitch1}. In addition, we explicitly present a series of attacks which are optimal with respect to this bound, closing the gap between the best known attacks and the tightest known bound. The article is organized as follows: after introducing notations in section~\ref{section:notations}, the new bounds are derived in section~\ref{section:iwasawa}. In section~\ref{section:homodyne}, we extend the technique used to establish these bounds to obtain new results about the homodyne protocols. Specifically, we will show that any homodyne attack using quantum memory is optimal, and that in some cases this optimality can be reached without quantum memory. Section~\ref{section:covariance} is devoted to another technique, based on symplectic invariants, which allows us to derive again the new heterodyne bounds from another point of view. Then section~\ref{section:attacks} describes four optimal attacks with respect to the new heterodyne bound, and section~\ref{section:discussion} concludes our study by discussing practical advantages of the heterodyne protocol.

\section{Notations}
\label{section:notations}

	In the case of Gaussian attacks, the channel linking Alice to Bob is fully characterized by its transmission in intensity $T$ (possibly greater than 1 for amplifying channels), and its excess noise $\epsilon$ above the shot noise level~\cite{pra}, such that the total noise measured by Bob is $(1 + T\epsilon)N_0$, where $N_0$ is the shot noise variance appearing in the Heisenberg relation $\mean{X^2} \mean{P^2} \geq N_0^2$. Alternatively we will make use of the total added noise referred to the input $\chi$ defined by $\chi=1/T+\epsilon-1$. These parameters might depend on the quadrature considered, in which case we will add a subscript indicating this quadrature (\emph{e.g.} $\chis{P}$).

	The quantum channel is considered to be probed by Eve with the help of ancillary modes. To index these modes, we will note $X_M$ and $P_M$ the quadratures of mode $M$, and write down the $2n$ quadratures of $n$ modes $M_1,\ldots,M_n$ by the vector $\mathbf{Q} = (X_{M_1},\ldots,X_{M_n},P_{M_1},\ldots,P_{M_n})$. The total Gaussian state of the system is then represented by its covariance matrix $\gamma$ of components $\gamma_{i,j} = \mean{\mathbf{Q}_i \mathbf{Q}_j}$.

	In the heterodyne protocol, we note $B$ the modes measured by Bob, and $B'$ the incoming beam, on which we will base our demonstrations. This mode is coupled with two ancillas on which Eve respectively measures $X$ and $P$ (Fig.~\ref{fig:heterogeneral}). In~\cite{Weedbrook:noswitch1}, the authors bounded the conditional variance $V_{X_{B'}|X_{E_1}}$ and $V_{P_{B'}|P_{E_2}}$ of the mode $B'$ knowing Eve's measurement results by
\begin{equation}
	\label{heterobound1}
	V^\minimum_{B'|E} = \frac{V}{T(1+ \chi V)}N_0,
\end{equation}
where $(V-1)N_0$ is Alice's modulation variance. This is basically the \emph{homodyne} RR bound derived in~\cite{Fred:nature,Fred:QIC} applied to each quadrature $X$ and $P$.

	Assuming that Eve does not control the beam-splitter $T_B$, this bound then leads to the minimal conditional variance of mode $B$ knowing Eve's measurement by adding the shot noise unit and the intensity decrease introduced by the beam-splitter $T_B$:
\begin{equation}
	V_{B|E} = \frac 1 2 \left(V_{B'|E} + N_0\right).
\end{equation}
Then the information acquired by Eve results from $V_{B|E}$:
\begin{equation*}
	I_{BE} = 2\times\frac 1 2 \log_2\left(\frac{V_B}{V_{B|E}}\right) \ \textrm{with}\ V_B= \frac{T(V + \chi) + 1}{2}N_0,
\end{equation*}
where the factor 2 reflects the double quadrature measurement. A similar reasoning on $I_{AB}$ finally gives the secret rate $\Delta I = I_{AB} - I_{BE}$. This rate is shown to be greater than the homodyne rate for any channel parameter, giving advantage to the heterodyne detection scheme.

\begin{figure}
\begin{center}
	\includegraphics[width=0.8\columnwidth]{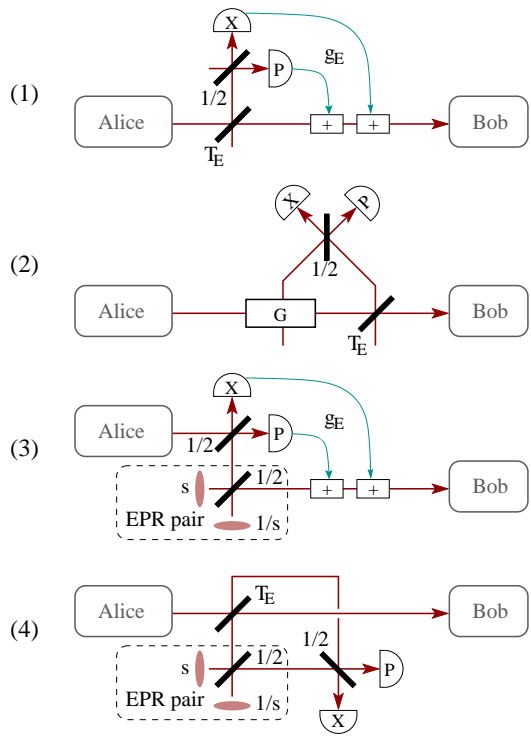}
	\caption{Our general results are illustrated by considering four optimal attacks against the heterodyne protocol. In the feed-forward attack (1), Eve taps a fraction $1-T_E$ of the signal on which she makes an heterodyne measurement. Then she translates Bob's quadratures according to her measurement results modified by a gain $g_E$. In the cloning attack (2), Eve amplifies the signal sent by Alice with a phase insensitive amplifier, and taps the amplified beam. Quantum teleportation (3) consist in a making the incoming beam interfere with a part of an EPR pair. $X$ and $P$ are measured in the output arms of the interferometer. Bob's quadratures are then translated according to Eve's results. Finally, in the entangling cloner attack (4), Eve tap a fraction $1-T_E$ of the incoming signal, while introducing some excess noise with the help of an EPR pair. The joint measurement of the tapped signal and the other part of the EPR pair optimally exploits both tapped signal and EPR noise correlations.}
	\label{fig:attaques}
\end{center}
\end{figure}

	Several explicit attacks against the heterodyne protocol have been considered. In~\cite{Weedbrook:noswitch1}, the authors propose an eavesdropping strategy based on heterodyne measurement and feed-forward (Fig.~\ref{fig:attaques}-1), which they numerically show to be suboptimal with respect to bound~(\ref{heterobound1}). In~\cite{Namiki}, Namiki \emph{et al.} introduce an eavesdropping strategy against the homodyne RRCVQKD protocol based on a cloning machine (Fig.~\ref{fig:attaques}-2). In this attack, Eve can measure both $X$ and $P$ quadratures of each coherent state, then requiring no quantum memory. The price to pay is that this attack is no more optimal with respect to the homodyne bound~(\ref{heterobound1}). In fact, the search for quantum-memory-less homodyne attacks is very similar to finding attacks on the heterodyne protocol because both schemes require that Eve measures $X$ and $P$ on each channel symbol. Therefore, the cloning attack is yet another sub-optimal attack against the heterodyne protocol when considering bound~(\ref{heterobound1}). We shall prove in this article that these two attacks are optimal with respect to the new bounds we derive for the heterodyne protocol, as well as two other attacks based on EPR entanglement.

\section{New bounds based on the Iwasawa symplectic decomposition}
\label{section:iwasawa}

	To derive the new bounds on heterodyne detection protocols, we will use the symplectic formalism which describes all physically possible Gaussian individual transformations on a set of $n$ modes. The real symplectic group is defined by the set of linear transformations of the quadrature vector $\mathbf{Q}$, which $2n \times 2n$ matrix $S$ satisfies
\begin{equation}
	\label{eqn:sympcond}
	S \beta \transpose{S} = \beta, \quad \textrm{with}\ \beta = \left[ \begin{array}{cc} 0 & \mathbb{I}_n \\ - \mathbb{I}_n & 0\end{array} \right],
\end{equation}
where $\mathbb{I}_n$ is the $n\times n$ identity matrix. The main idea of our demonstration is to use a proper parameterization that spans all symplectic transformations applied to the modes going through the quantum channel, hence all possible attacks, and to compute the best information Eve can obtain when these parameters vary.

	The real symplectic group is a $n(2n+1)$ parameters space for which various parameterizations -- or decompositions -- exist. We choose the Iwasawa decomposition~\cite{Simon}, which uniquely factorizes any $2n\times 2n$ symplectic matrix $S$ as the product of 3 special matrices:
\begin{equation}
	S = \left[\begin{array}{cc} A & 0 \\ C & \transpose{\left(A^{-1}\right)}\end{array} \right]
	    \left[\begin{array}{cc} D & 0 \\ 0 & D^{-1}\end{array} \right]
	    \left[\begin{array}{cc} B & F \\ -F & B\end{array} \right]
\end{equation}
where $B+iF$ is a $n\times n$ unitary matrix, $D$ is diagonal with strictly positive components, $A$ is lower triangular with all diagonal terms set to 1, and $\transpose{A}C$ is symmetric.

	In our study, we consider 3 modes, depicted in Fig.~\ref{fig:heterogeneral}. The first one, noted $B'$ is send from Alice to Bob, who performs an heterodyne measurement upon reception. Eve makes this mode interact with two ancillary modes $E_1$ and $E_2$ initially in the vacuum state, and then measures $X$ on $E_1$, and $P$ on $E_2$. To respect the symmetry of this problem, we only consider symplectic transformations $S$ that do not mix $X$ and $P$ quadratures:
\begin{equation}
	\label{eqn:blockdiag}
	S = \left[\begin{array}{cc} S_X & 0 \\ 0 & S_P\end{array} \right].
\end{equation}
As $B$, $D$ and $A$ are invertible, expanding $S$ yields $F = C = 0$, and $B$ orthogonal. We recall that the columns of the orthogonal matrix $B$, as well as its rows, form an orthonormal basis. With this form, the Iwasawa decomposition has a physical meaning in terms of linear optical components. Namely, any symplectic transformation is composed of an orthogonal transformation which is itself a composition of rotations (\emph{i.e.} beam-splitters) and reflections (\emph{i.e.} $\pi$-phase shifts), 1-mode squeezers, and feed-forward.

Finally, we can write the Iwasawa decomposition corresponding to our attack model:
\begin{eqnarray}
	S_X & = & \left[\begin{array}{ccc}
			1 & 0 & 0 \\
			a & 1 & 0 \\
			b & c & 1 \\
		\end{array}\right]
	\left[\begin{array}{ccc}
			s_1 & 0 & 0 \\
			0 & s_2 & 0 \\
			0 & 0 & s_3 \\
		\end{array}\right]
		\left[\begin{array}{ccc}
			b_1 & b_2 & b_3 \\
			b_4 & b_5 & b_6 \\
			b_7 & b_8 & b_9 \\
		\end{array}\right]\\
	\nonumber
	S_P & = & \left[\begin{array}{ccc}
			1 & -a & \delta \\
			0 & 1 & -c \\
			0 & 0 & 1 \\
		\end{array}\right]
		\left[\begin{array}{ccc}
			s_1^{-1} & 0 & 0 \\
			0 & s_2^{-1} & 0 \\
			0 & 0 & s_3^{-1} \\
		\end{array}\right]
		\left[\begin{array}{ccc}
			b_1 & b_2 & b_3 \\
			b_4 & b_5 & b_6 \\
			b_7 & b_8 & b_9 \\
		\end{array}\right],
\end{eqnarray}
with $\delta = ac-b$. The orthogonal matrix $B$ can be parameterized by 3 real parameters (Euler angles) plus a binary parameter (the sign of the determinant). This leaves 9 real and one discrete symplectic parameters to characterize the matrix $S$. By expanding $S$ and using orthogonality properties of $B$, we can express channel parameters as functions of these symplectic parameters:
\begin{eqnarray}
	\nonumber
	T_X & = & t_X^2 \ \textrm{with} \ t_X = S_{1,1} = s_1 b_1 \\
	\label{chanparam}
	T_P & = & t_P^2 \ \textrm{with} \ t_P = S_{4,4} = \frac{b_1}{s_1} - \frac{a b_4}{s_2} + \frac{\delta b_7}{s_3} \\
	\nonumber
	\chis{X} & = & \frac{{S_{1,2}}^2 + {S_{1,3}}^2}{T_X} = \frac{1}{b_1^2}-1 \\
	\nonumber
	\chis{P} & = & \frac{{S_{4,5}}^2 + {S_{4,6}}^2}{T_P} = \left[\frac{1}{s_1^2} + \frac{a^2}{s_2^2} +
	\frac{\delta^2}{s_3^2}\right]\frac{1}{T_P}-1
\end{eqnarray}

We note that $s_1$ and $b_1$ are equivalent to channel parameters $T_X$ and $\chis{X}$. As we are looking for the best attack for given channel parameters, we will consider $s_1$ and $b_1$ fixed. Our attacks are then characterized by 7 real and one discrete parameters.

The input covariance matrix $\gamma^i$ is diagonal with diagonal terms $(V,1,1,V,1,1)N_0$. The output covariance matrix is expressed as $\gamma = S \gamma^i \transpose{S}$. From $S$ and $\gamma$, we obtain Eve's noises and conditional variances
\begin{eqnarray}
	\nonumber
	\chis{X_{E_1}} & = & \frac{{S_{2,2}}^2 + {S_{2,3}}^2}{{S_{2,1}}^2} = \frac{r^2 + 1}{(r b_1 + b_4)^2} - 1 \\
	\nonumber
	\chis{P_{E_2}} & = & \frac{{S_{6,5}}^2 + {S_{6,6}}^2}{{S_{6,4}}^2} = \frac{1}{1 - b_1^2 - b_4^2} - 1 \\
	\label{eveparam}
	\frac{V_{X_{B'}|X_{E_1}}}{N_0}
		& = &
		\gamma_{1,1} - \frac{\gamma_{1,2}\gamma_{2,1}}{\gamma_{2,2}} \\
		\nonumber
		& = & \frac{s_1^2}{r^2+1}\frac{(V\chis{P_{E_2}}+1)(\chis{X_{E_1}}+1)}{(V+\chis{X_{E_1}})(\chis{P_{E_2}}+1)}\\
	\nonumber
	\frac{V_{P_{B'}|P_{E_2}}}{N_0}
		& = &
		\gamma_{4,4} - \frac{\gamma_{4,6}\gamma_{6,4}}{\gamma_{6,6}} \\
		\nonumber
		& = & \frac{r^2+1}{s_1^2}\frac{(V\chis{X_{E_1}}+1)(\chis{P_{E_2}}+1)}{(V+\chis{P_{E_2}})(\chis{X_{E_1}}+1)}
\end{eqnarray}
where $r = a s_1/s_2$. All these quantities only depend on parameters $r$ and $b_4$: our parameter space drops to 2 parameters.

	Then, we will require that the attack leaves channel parameters symmetric in $X$ and $P$, \emph{i.e.} $T_X = T_P \equiv T$ and $\chis{X} = \chis{P} \equiv \chi$. The former relation univocally fixes $\delta = s_3(b_1(s_1^2-1)/s_1+ab_4/s_2)/b_7$, and the later fixes $r$:
\begin{equation}
	r = \frac{b_1 b_4(1-s_1^2) + \sigma \sqrt{(1-b_1^2-b_4^2)\rho}}{1-b_1^2}.
\end{equation}
where $\sigma = \pm 1$ and $\rho = (s_1^2-1)(1-s_1^2(2b_1^2-1))$. In terms of channel parameters, $\rho = (T\chi)^2 - (1-T)^2$. For a symmetric channel, the Heisenberg inequality requires that $\rho \geq 0$~\cite{caves}, therefore $r$ is well defined for any attack that can be made symmetric. Finally, we are left with only one parameter, $b_4$, such that $b_4^2 < 1-b_1^2 = \chi/(1+\chi)$, and the sign $\sigma$.

	For RR, the information Eve acquires is given by
\begin{equation}
	\nonumber
	I_{BE} = \frac 1 2 \log_2\left(\frac{V_B}{V_{X_B|X_{E_1}}}\right) + \frac 1 2 \log_2\left(\frac{V_B}{V_{P_B|P_{E_2}}}\right).
\end{equation}
$I_{BE}$ is maximum when $(V_{X_{B'}|X_{E_1}} + 1)(V_{P_{B'}|P_{E_2}} + 1)$ is minimum. For direct reconciliation (DR), for which the key is distilled from Alice's data, the information Eve acquires is given by the Shannon formula:
\begin{equation}
	\nonumber
	I_{AE} = \frac 1 2 \log_2\left(\frac{V + \chis{X_{E_1}}}{1+\chis{X_{E_1}}}\right) + \frac 1 2 \log_2\left(\frac{V + \chis{P_{E_2}}}{1+\chis{P_{E_2}}}\right).
\end{equation}

We find that both mutual informations $I_{AE}$ and $I_{BE}$ have an extremum at
\begin{equation}
	b_4 = \sigma \frac{s_1 \sqrt{1-s_1^2(2b_1^2 -1)} - \sqrt{b_1^2(s_1^2-1)}}{s_1^2+1}
\end{equation}
We check numerically that in the quantum regime defined by $\epsilon \leq 2$, this extremum is indeed the absolute maximum. For this value of $b_4$, we compute Eve's noise and conditional variance as functions of channel parameters:
\begin{equation}
	\label{heterobound3}
	\chis{X_{E_1}} = \chis{P_{E_2}} \equiv \chis{E}^\minimum = \frac{T(2-\epsilon)^2}{(\sqrt{2-2T+T\epsilon} + \sqrt{\epsilon})^2} + 1\\
\end{equation}
\begin{equation}
	\label{heterobound4}
	V_{X_{B'}|X_{E_1}} = V_{P_{B'}|P_{E_2}} \equiv V_{B'|E}^\minimum = \frac{V\chis{E} + 1}{V + \chis{E}}N_0
\end{equation}
These expressions form the new bounds for direct and reverse reconciliated heterodyne protocols. As they are obtained for the same value of $b_4$, any attack that reaches bound~(\ref{heterobound3}) (\emph{i.e.} optimal for DR) also reaches bound~(\ref{heterobound4}) (\emph{i.e.} optimal for RR).

\section{Application to the homodyne detection protocol}
\label{section:homodyne}

	In this section, we will show that bound~(\ref{heterobound1}) on the homodyne protocol can also be derived from the Iwasawa symplectic decomposition. In the homodyne protocol, Eve stores the quantum states of mode $E_1$ and mode $E_2$ in quantum memories, waiting for Bob's measurement basis disclosure. After this, Eve can measure the \emph{same} quadrature $Q = X$ or $P$ chosen by Bob on \emph{both} modes. The information acquired by Eve in the RR homodyne protocol is deduced from the conditional variance on Bob's measurement knowing the quadrature $Q$ of modes $E_1$ and $E_2$, which can be computed from the output covariance matrix $\gamma$:
\begin{equation}
	V_{Q_{B'}|Q_{E_1}, Q_{E_2}} = \frac{\det(\gamma_Q)}{\det(\gamma_E)},
\end{equation}
where $\gamma_Q$ is the restriction of $\gamma$ to the quadrature $Q$, and $\gamma_E$ is the restriction of $\gamma$ to the quadrature $Q$ of Eve's modes $E_1$ and $E_2$. By expanding the Iwasawa decomposition of the symplectic transformation $S$ decribing the attack, and by using orthogonality properties of matrix $B$, we can express this conditional variance as:
\begin{equation}
	\label{homcondvar}
	V_{Q_{B'}|Q_{E_1}, Q_{E_2}} = \frac{V}{T_{Q'}(V\chis{Q'}+1)}N_0,
\end{equation}
	where $Q' = P$ or $X$ is the quadrature \emph{not} measured by Bob. This conditional variance coincides with the homodyne bound~(\ref{heterobound1}). It is important to note that contrary to the heterodyne conditional variance which depends on symplectic parameters $r$ and $b_4$ as shown by equations~(\ref{eveparam}), the homodyne conditional variance~(\ref{homcondvar}) only depends on channel parameters, but no other symplectic parameter characterizing the attack

	The DR case is treated similarly, by considering the covariance matrix $\gamma^{AE}$ that gathers the modulation value chosen by Alice $(X_A,P_A)$ and modes $E_1$ and $E_2$ owned by Eve. By expanding the Iwasawa decomposition of $S$, we find
\begin{equation}
	\label{homcondvarDR}
	V_{Q_A|Q_{E_1}, Q_{E_2}} = \frac{\det(\gamma_Q^{AE})}{\det(\gamma_E)} = \frac{(V-1)(1+\chis{Q'})}{V\chis{Q'}+1}N_0,
\end{equation}
which yields
\begin{equation*}
	I_{AE} = \frac 1 2 \log_2\left(\frac{(V-1)N_0}{V_{Q_A|Q_{E_1}, Q_{E_2}}}\right) = \frac 1 2 \log_2\left(\frac{V+\chis{E}^\textrm{hom}}{1+\chis{E}^\textrm{hom}}\right),
\end{equation*}
with $\chis{E}^\textrm{hom} = 1/\chis{Q'}$. This expression matches the highest bound for the information acquired by Eve in the DR homodyne protocol established in~\cite{Fred:DR}. It depends only on channel parameters, and not on the other symplectic parameters.

	Therefore, we have shown any attack against the DR or RR homodyne protocols
\begin{enumerate}
	\item \label{cond:mix} that do not mix quadratures $X$ and $P$
	\item \label{cond:two} that can be performed with two ancillary modes
	\item \label{cond:vac} in which all ancillas are initially vacuum states
	\item \label{cond:all} in which Eve measures the same quadrature as Bob on all of her ancillary modes with the help of quantum memories
\end{enumerate}
is optimal for the channel parameters it can reproduce. In particular, the entangling cloner attack introduced in~\cite{Fred:nature} and the assymetric cloning attack studied in~\cite{Namiki} are optimal homodyne attacks~; for channels with no excess noise ($\epsilon = 0$), the beam-splitting attack is optimal. Equations~(\ref{homcondvar}) and~(\ref{homcondvarDR}) show that the optimality of any attack that verify conditions 1--4 holds even for attacks yielding dissymmetric channel parameters (\emph{i.e.} $T_X \neq T_P$ and $\chis{X} \neq \chis{P}$).

	In fact, conditions 1--3 do not hamper the generality of the attacks we consider. Indeed, condition~\ref{cond:mix}. is not restrictive as one can reduce any symplectic matrix to the block diagonal form~(\ref{eqn:blockdiag}) by means of local Gaussian operations~\cite{Duan}. We will see in section \ref{section:discussion} that numerical simulations show that condition~\ref{cond:two}. is in fact not necessary. Finally, condition~\ref{cond:vac}. is also not restrictive because one can include the preparation of a non-vacuum initial state from vacuum states inside the symplectic transformation describing the attack, eventually by making use of extra ancillary modes.

	The fact that the heterodyne attack scheme breaks condition~\ref{cond:all}. is the reason why, in general, the heterodyne bound~(\ref{heterobound4}) is higher than the homodyne bound~(\ref{heterobound1}), thus imposing more stringent constraints on Eve's information. However, for some particular values of the channel parameters, these two bounds coincide. Namely, this happens when $\chi = \sqrt{1-T+T/V^2}/T - 1/V$ for RR. For DR, bound~(\ref{heterobound3}) is equal to its homodyne counterpart $\chis{E}^\textrm{hom} = 1/\chi$ when $\chi = \sqrt{1-1/T}$ with $T\geq 1$. For these channel parameters, an optimal heterodyne attack is also an optimal homodyne attack, but without the need for quantum memories, therefore lowering the technological requirements for the eavesdropper.

	We recall that like all the results presented in this paper, the optimality of any homodyne attack is to be understood in the context of individual Gaussian attacks. However, since the homodyne bound~(\ref{heterobound1}) is proven secure against the larger class of individual and finite-size Gaussian and non-Gaussian attacks~\cite{Fred:ng}, we can say that any Gaussian individual attack that fulfills conditions 1--4 is optimal among that extended class of attacks. Security proofs of the homodyne protocol against collective attacks require the use of the Holevo entropy~\cite{Fred:col,Raul:col}, then the results presented here do not apply to this general class of attacks.

\section{Proof based on symplectic invariants}
\label{section:covariance}

	It is possible to derive the heterodyne bound~(\ref{heterobound4}) from another technique that does not require the Iwasawa decomposition. This technique is based on the fact that the output covariance matrix $\gamma$ issues from some symplectic transformation $S$ applied to the initial covariance matrix $\gamma^i$. Since $\gamma^i$ is diagonal with diagonal terms $(V,1,1,V,1,1)N_0$, this property simply states that $(V,1,1)N_0$ are the symplectic eigenvalues of the output covariance matrix $\gamma$. In other word, finding the best attack for RR amounts to minimizing the conditional variance of Bob's measurement knowing Eve's measurement over the set of covariance matrices with symplectic eigenvalues $(V,1,1)N_0$. In terms of symplectic transformations, Heisenberg relations on the three modes we consider require that all symplectic eigenvalues are greater than $N_0$. Therefore, covariance matrices with eigenvalues $(V,1,1)N_0$ are covariance matrices that are compatible with Heisenberg relations and an input modulation of variance $VN_0$.

	Since symplectic eigenvalues are usually hard to express analytically, we will rather use symplectic invariants, which are totally equivalent to symplectic eigenvalues. For a three mode state, there exist three symplectic invariants $\Delta_{j,3}$ with $j=1,2,3$ defined as the sum of the determinant of all $2j\times 2j$ sub-matrices of $\gamma$ which diagonal is on the diagonal of $\gamma$~\cite{adesso}. Applied to the input covariance matrix $\gamma^i$, these invariants read
\begin{eqnarray}
	\Delta_{1,3} & = & V^2 + 2\\
	\Delta_{2,3} & = & 2V^2 + 1\\
	\Delta_{3,3} & = & V^2
\end{eqnarray}
We will now express the symplectic invariants as functions of the components of the output covariance matrix $\gamma$. For this purpose, we write this matrix as
\begin{equation}
	\label{gammaout}
	\gamma =
	\left[
	\begin{array}{llllll}
		V_{B'} & c_m & c_n & 0 & 0 & 0 \\
		c_m & V_{Em} & c & 0 & 0 & 0 \\
		c_n & c & V_{En} & 0 & 0 & 0 \\
		0 & 0 & 0 & V_{B'} & c_n & c_m \\
		0 & 0 & 0 & c_n & V_{En} & c \\
		0 & 0 & 0 & c_m & c & V_{Em}
	\end{array}
	\right]N_0
\end{equation}
where $m$ stands for ``measured" and $n$ for ``not measured", and $V_{B'} = T(V+\chi)$. This notation assumes that the attack does not mix quadratures $X$ and $P$, and that swapping measurements of modes $E_1$ and $E_2$ would not change Eve's information. The later assumption is backed by results of section~\ref{section:iwasawa} where we found that the optimal heterodyne attack yields to equal variances for $X$ and $P$ measurements. From equation~(\ref{gammaout}) we can compute symplectic invariants as
\begin{eqnarray*}
	\Delta_{1,3} & = & 2c^2 + V_{B'}^2 + 4x + 2y\\
	\Delta_{2,3} & = & c^4 + 2c^2V_{B'}^2 + 4c^2x - 4cV_{B'}x + 4x^2 - 2c^2y \\
					&   & + 2V_{B'}^2y + 4xy + y^2 - 4cz - 2V_{B'}z\\
	\Delta_{3,3} & = & (-2cx + V_{B'}(c^2-y) + z)^2,
\end{eqnarray*}
where we introduced variables
\begin{equation*}
	x = c_m c_n,\quad y = V_{Em} V_{En}, \quad z = V_{Em}c_n^2 + V_{En}c_m^2.
\end{equation*}
With these variables, Eve's conditional variance yields
\begin{eqnarray*}
	V_{B'|E} & = & \left( V_{B'} - \frac{c_m^2}{V_{Em}} \right)N_0 \\
	         & = & \left( V_{B'} - \frac{z+\sigma'\sqrt{z^2-4yx^2}}{2y} \right)N_0,\ \sigma' = \pm 1
\end{eqnarray*}
Since $y > 0$, we will only consider $\sigma' = 1$ because it gives more information to Eve. Using the invariance of symplectic invariants, we univocally fix $x$ and $z$
\begin{eqnarray*}
	x & = & \frac 1 4 \left(2(1-c^2-y)+V^2-V_{B'}^2\right) \\
	z & = & V_{B'}y - V - c^2(V_{B'}+c) + \frac c 2 \left[2(1-y) + V^2 - V_{B'}^2\right],
\end{eqnarray*}
as well as $c$, as a function of channel parameters
\begin{equation}
    c = \frac{V-V_{B'}}{2}.
\end{equation}
Consequently, the heterodyne conditional variance $V_{B'|E}$ only depends on channel parameters and $y$. Then, we notice that $y$ appears in the homodyne conditional variance
\begin{equation}
	V_{B'|E_1,E_2} = \frac{V}{y-c^2} N_0.
\end{equation}
Therefore, the homodyne bound~(\ref{heterobound1}) contraints $y$ by
\begin{equation}
    c^2 \leq y \leq c^2 + T(V\chi + 1)
\end{equation}
We numerically find that $V_{B'|E}$ is a decreasing function of $y$, therefore the highest value for $y$ must be considered to bound Eve's information. In fact, using the results of section~\ref{section:homodyne} stating that any attack on the homodyne protocole is optimal (the covariance matrix~(\ref{gammaout}) fulfills conditions 1--4), we can say that the only possible value for $y$ is indeed $c^2 + T(V\chi + 1)$. Now, $V_{B'|E}$ only depends on channel parameters, and we can check that it coincides with bound~(\ref{heterobound4}).

	In conclusion, we have shown another technique for deriving bound~(\ref{heterobound4}). This technique is slightly less general than the Iwasawa decomposition because it assumes that the optimal attack respects the symmetry of the problem. Furthermore, it does not cover the DR protocol. Yet, it enables to find bound~(\ref{heterobound4}) from more fundamental Heisenberg-like properties.

\section{Optimal attacks}
\label{section:attacks}

	We shall now exhibit four optimal attacks against the heterodyne protocol with repect to bounds~(\ref{heterobound3}) and~(\ref{heterobound4}), depicted in Fig.~\ref{fig:attaques}. The existence of such optimal attacks show that the bounds we derived are tight: it is not possible to further reduce the estimation Alice and Bob can make about Eve's information. The first optimal attack we consider is the feed forward attack introduced in~\cite{Weedbrook:noswitch1}. The symplectic matrix associated with this attack is
\begin{equation*}
	S_X^{ff} = \left[\begin{array}{ccc}
			1 & g_E & 0 \\
			0 & 1 & 0 \\
			0 & 0 & 1 \\
		\end{array}\right]
		\left[\begin{array}{ccc}
			1 & 0 & 0 \\
			0 & \frac{1}{\sqrt 2} & \frac{1}{\sqrt 2} \\
			0 & \frac{-1}{\sqrt 2} & \frac{1}{\sqrt 2} \\
		\end{array}\right]
		\left[\begin{array}{ccc}
			\sqrt{\scriptscriptstyle T_E} & -\sqrt{\scriptscriptstyle 1-T_E} & 0 \\
			\sqrt{\scriptscriptstyle 1-T_E} & \sqrt{\scriptscriptstyle T_E} & 0 \\
			0 & 0 & 1 \\
		\end{array}\right]
\end{equation*}
and $S_P^{ff}$ is obtained from $S_X^{ff}$ by replacing the first line in the leftmost matrix by $[1,0,-g_E]$. Using equations~(\ref{chanparam}) which link coefficients of the symplectic transformation matrix to channel parameters $T$ and $\epsilon$, we can see that to faithfully reproduce these channel parameters, Eve must choose:
\begin{equation*}
	g_E^2 = \epsilon T, \quad
	T_E = 4\frac{2 - \sqrt{\epsilon(2-2T+T\epsilon))}}{(2+T\epsilon)^2 /T} - T\frac{(2-\epsilon)}{(2+T\epsilon)}
\end{equation*}
With these parameters, we can check from the components of $S_X^{ff}$ injected in equations~(\ref{eveparam}) that this attack reaches bounds~(\ref{heterobound3}) and~(\ref{heterobound4}).

Quantum teleportation is represented by the symplectic matrix
\begin{eqnarray*}
	S_X^{qt} & = & \left[\begin{array}{ccc}
			1 & g_E & 0 \\
			0 & 1 & 0 \\
			0 & 0 & 1 \\
		\end{array}\right]
		\left[\begin{array}{ccc}
			0 & 0 & 1 \\
			0 & 1 & 0 \\
			1 & 0 & 0 \\
		\end{array}\right]
		\left[\begin{array}{ccc}
			\frac{1}{\sqrt 2} & \frac{-1}{\sqrt 2} & 0\\
			\frac{1}{\sqrt 2} & \frac{1}{\sqrt 2} & 0\\
			0 & 0 & 1 \\
		\end{array}\right] S_X^{EPR} \\
		\textrm{with} & & S_X^{EPR} = \left[\begin{array}{ccc}
			1 & 0 & 0 \\
			0 & \frac{1}{\sqrt 2} & \frac{-1}{\sqrt 2} \\
			0 & \frac{1}{\sqrt 2} & \frac{1}{\sqrt 2} \\
		\end{array}\right]
		\left[\begin{array}{ccc}
			1 & 0 & 0 \\
			0 & s^{-1} & 0 \\
			0 & 0 & s \\
		\end{array}\right]
\end{eqnarray*}
Here, $S_P^{qt}$ is obtained by using $[1,0,g_E]$ as the first line of the leftmost matrix, and changing $s\rightarrow 1/s$. The second from left matrix simply swaps the 1st and 3rd modes to respect our mode order convention. Channel parameters fix $s$ and $g_E$:
\begin{equation*}
	g_E^2 = 2T, \quad s^2 = \frac{1-T+T\epsilon-\sqrt{T\epsilon(2-2T+T\epsilon)}}{(1-\sqrt{T})^2}
\end{equation*}
and noise computation shows that this attack is optimal.

	Then, the entangling cloner attack is represented by the symplectic matrix
\begin{equation*}
	S_X^{ec} = \left[\begin{array}{ccc}
			1 & 0 & 0 \\
			0 & \frac{1}{\sqrt 2} & \frac{-1}{\sqrt 2}\\
			0 & \frac{1}{\sqrt 2} & \frac{1}{\sqrt 2}\\
		\end{array}\right]
		\left[\begin{array}{ccc}
			\sqrt{\scriptscriptstyle T_E} & -\sqrt{\scriptscriptstyle 1-T_E} & 0 \\
			\sqrt{\scriptscriptstyle 1-T_E} & \sqrt{\scriptscriptstyle T_E} & 0 \\
			0 & 0 & 1 \\
		\end{array}\right]
		S_X^{EPR}
\end{equation*}
To fake channel parameters, Eve must choose
\begin{equation*}
	T_E = T \quad \textrm{and} \quad \frac{s^4+1}{2s^2} = \frac{T\epsilon}{1-T} + 1
\end{equation*}
Once again, noise and conditional variance computations from the components of this matrix yield bounds~(\ref{heterobound3}) and~(\ref{heterobound4}).

	Finally, the cloning attack studied in~\cite{Namiki} is also optimal. This can be checked by verifying that the conditional variance of equation~(44) in~\cite{Namiki} coincides with bound~(\ref{heterobound4}). For this attacks, the authors show that in order to reproduce channel parameters, Eve must choose
\begin{equation*}
	T_E = T(1-\epsilon/2) \quad \textrm{and} \quad G = \frac{1}{1-\epsilon/2}.
\end{equation*}

	For no excess noise ($\epsilon = 0$), all these attacks are equivalent to beam-splitting attacks.

\begin{figure}
\begin{center}
	\includegraphics[width=\columnwidth]{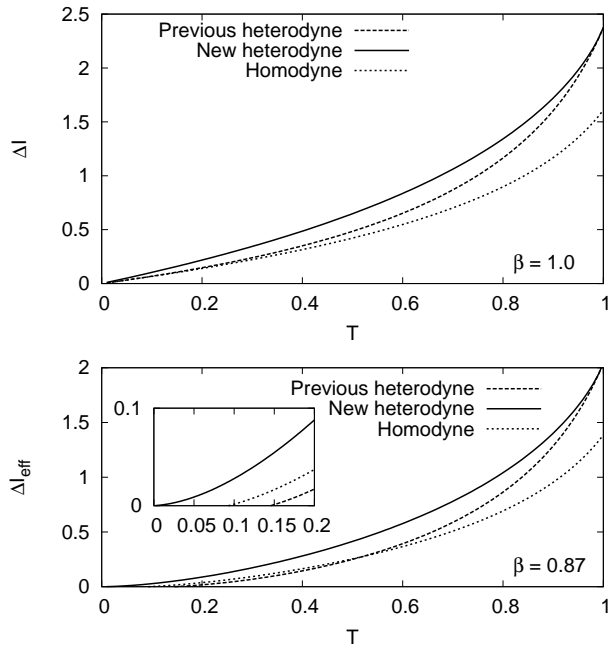}
	\caption{Effective information rate for typical experimental parameters: $V = 11$, $\epsilon=0.02$ and a perfect error correction $\beta = 1$ (top) or a constant reconciliation efficiency $\beta = 0.87$ (bottom). The new bound on heterodyne protocol provides more secret information than the homodyne protocol or the previous heterodyne bound. We can see from the bottom graph inlet that with these parameters, the new heterodyne provides secret information for every channel transmission. However, in practice, the reconciliation efficiency $\beta$ drops as the distance rises, then limiting the range of the protocol.}
	\label{fig:info}
\end{center}
\end{figure}

\section{Discussion}
\label{section:discussion}

	We first discuss the generality of the model shown on Fig.~\ref{fig:heterogeneral}, on which we built our proofs. This model assumes that Eve's attack only involves two modes, but one can imagine that Eve could use and measure more modes, also carrying some information about Alice and Bob's transmission. To tackle this problem, we can imagine that Eve concentrates all the information her modes bear into a single mode for each quadratures, by iterative constructive interferences between her modes using beam-splitters. Since local operations using beam-splitters on Eve's modes do not alter the conditional variance $V_{B'|E}$, any attack on $n$ modes for each quadrature $X$ and $P$ can be mapped to an equivalent attack, where Eve only needs to measure one mode for each quadrature. Therefore, it seems reasonable to assess that it is useless for Eve to introduce extra modes that in the end do not provide any information about Alice and Bob's data. This technique is illustrated in~\cite{Namiki}, where the authors consider the interference of the two modes owned by Eve in the assymetric attack against the homodyne protocol, and show that this interference enables Eve to measure only one mode without loosing information.

	To back this argument, we performed numerical simulations that give $2n$ modes to Eve, with $n=1,2,5$. In these simulations, $10^7$ attacks are tested by generating random symplectic transformations parameterized by the Iwasawa decomposition. It shows that the two main results of this paper hold with more that two modes for Eve, namely that any attack using quantum memories on the homodyne protocol is optimal, and that the information Eve can get on the heterodyne protocol is bounded by~(\ref{heterobound4}).

	We complete our study by discussing practical advantages of the heterodyne scheme over the homodyne scheme, when considering that a classical error correction with limited efficiency $\beta$ has to be applied to experimental data to obtain a secret key~\cite{bloch:LDPC}. In this picture, the practical key rate becomes $\Delta I_\textrm{eff} = \beta I_{AB} - I_{BE}$, resulting in a bit loss of $\Delta I - \Delta I_\textrm{eff} = (1-\beta)I_{AB}$. Because for a given efficiency $\beta$ the mutual information $I_{AB}$ of the heterodyne scheme is higher, this protocols suffers from greater key loss than the homodyne scheme. When considering bound~(\ref{heterobound1}), this loss was rapidly fatal. However, with the new bound~(\ref{heterobound4}), we can see from Fig.~\ref{fig:info} that the heterodyne scheme recovers its advantage.

	Still, there are two other practical drawbacks to the heterodyne protocol. First, for a given distance, the signal to noise ratio (SNR) of the transmission is lower because of Bob's heterodyning beam-splitter. Since the reconciliation efficiency is an increasing function of the SNR, this effect lowers the final key rate. Because of this, both heterodyne and homodyne protocols feature an equivalent key rate. For example, for $T = 0.25$ (corresponding to 25 km), $\epsilon \simeq 0,02$ and $V \simeq 11 N_0$~\footnote{A higher modulation variance $V$ could be used to increase the SNR, thus compensating for the SNR decrease due to Bob's heterodyne measurement. However, a higher modulation variance also increases the information $I_{BE}$. Then, the effective secret rate $\Delta I_\textrm{eff}$ features a maximum for a certain value of $V$, which turns out to be comparable for both heterodyne and homodyne protocols for the given channel parameters.}, the homodyne scheme achieves $\beta = 0.87$ and the heterodyne scheme $\beta = 0.80$, both yielding to a few 0.01 bits per symbol. Second, Alice and Bob need to reconcile twice as much data as for the homodyne case. This effect also lowers the final key rate when, as experimentally observed, computing speed limits the experimental repetition rate. However, on-going work on reconciliation at low SNR may take advantage of the high effective key rate of the heterodyne protocol.

	In conclusion, we have derived new bounds for individual attacks on the direct and reverse reconciliated QKD protocols with heterodyne detection. These new bounds offer a higher secret key rate than previous bounds. We have shown that the feed-forward attack, the cloning attack, the quantum teleportation and the entangling cloner all achieve these bounds, then closing the gap between best known bounds and best known attacks. On the other hand, the behaviour of these new bounds with respect to non-Gaussian~\cite{Fred:ng} and collective attacks~\cite{Fred:col,Raul:col} remains an open question.

	We thank Fr\'ed\'eric Grosshans, Raul Garc\'ia-Patr\'on and Nicolas Cerf for fruitfull discussions. We acknowledge support from the SECOQC European Integrated Project. J.L. acknowledges support from IFRAF.

\bibliography{hetero}

\end{document}